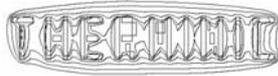

Nice, Côte d'Azur, France, 27-29 September 2006# CCD THERMOREFLECTANCE THERMOGRAPHY SYSTEM: METHODOLOGY AND EXPERIMENTAL VALIDATION

*Pavel L. Komarov, Mihai G. Burzo, and Peter E. Raad*

Nanoscale Electro-Thermal Sciences Laboratory
Department of Mechanical Engineering
Southern Methodist University
Dallas, TX 75275-0337, U.S.A**ABSTRACT**

This work introduces a thermoreflectance-based system designed to measure the surface temperature field of activated microelectronic devices at submicron spatial resolution with either a laser or a CCD camera. The article describes the system, outlines the measurement methodology, and presents validation results. The thermo-reflectance thermography (TRTG) system is capable of acquiring device surface temperature fields at up to 512×512 points with 0.2 μm resolution. The setup and measurement methodology are presented, along with details of the calibration process required to convert changes in measured surface reflectivity to absolute temperatures. To demonstrate the system's capabilities, standard gold micro-resistors are activated and their surface temperature fields are measured. The results of the CCD camera and our existing laser-based measurement approaches are compared and found to be in very good agreement. Finally, the system is validated by comparing the temperatures obtained with the TRTG method with those obtained from electrical resistance measurements.## 1. INTRODUCTION

The high power and speed required by modern electronic devices have translated into dramatic increases in the density of elementary transistors coupled with equally dramatic decreases in feature sizes. As a result, there is a need for critical solutions to the problem of heat removal from high-power high-density elementary devices. The geometric and material complexities in leading edge electronic devices make predictions and simulations of thermal behavior quite difficult. Thus, additional direct tools are needed to test the real behavior of such electronic structures, if significant design improvements are to be enabled. As a result, there has been an increased demand for methods that can directly measure the temperature of the features of such structures, which more than often are at the submicron level [1-3]. Contact methods can be used, but they present the added difficulties of having to access features of a submicron device with an external probe, or in the case of embedded features, fabricate a measuring probe into the device, and then having to isolate and exclude the influence of the measuring probe itself [4]. Alternatively, optical methods can be used, among which, the thermoreflectance method possesses important advantages and is so far one of the methods that has been successfully employed to make submicron temperature mappings [1-3, 5-10]. Thermo-reflectance thermography (TRTG) is an efficient non-contact and non-destructive optical approach for probing steady-state and transient surface temperature, providing accurate results for submicron features of microelectronic devices with excellent spatial and thermal resolutions.

Thermoreflectance microscopy is based on the principle that a change in the temperature of a given material produces a small change in the reflectivity of that material's surface. Thus, to measure the increase (or decrease) in the temperature of a sample, $\Delta T$, one needs to measure the change in the reflectivity of the sample $\Delta R/R$ and the thermoreflectance calibration coefficient $C_{TR}$. As might be anticipated, the most challenging aspect for thermoreflectance measurements is the small value of the thermoreflectance coefficient of the top layer material, $C_{TR}$, which defines the rate of change in the surface reflectivity as a function of a change in surface temperature. The $C_{TR}$ coefficient needs to be sufficiently high in order to have an appropriate signal-to-noise ratio in the measurements. Usually, it must be higher than $10^{-5}$ per Kelvin in order to obtain a temperature distribution with a good level of accuracy. The primary factors that influence $C_{TR}$ are the material under test, the wavelength of the probing laser [11-14], and the composition of the sample (if multi-layered) [10, 14, 15].

The goals of this work are to show that: (1) the developed system is capable of detecting the temperature increases of an activated elementary device (a gold resistor was chosen); (2) the obtained results compare favorably with results obtained by the use a single-point

©TIMA Editions/THERMINIC 2006    -page-    ISBN: 2-916187-04-9

*Pavel L. Komarov, Mihai G. Burzo, and Peter E. Raad*
*CCD Thermoreflectance Thermography: Methodology and Experimental Validation*

laser-based approach, while revealing the advantages of the multipoint CCD camera approach; and (3) the measured changes are consistent with the measurement results obtained by the use of a method based on non-optical physics, namely, the thermo-electrical resistance.

## 2. MEASUREMENT METHODOLOGY

The experimental thermography temperature system is based on the thermoreflectance (TR) method, where the change in surface temperature is measured by detecting the change in the reflectivity of the sample. Since the change in reflectivity (i.e., thermo-reflectance coefficient, $C_{TR}$) is of the order of $10^{-3}$–$10^{-5}$ K$^{-1}$ for most electronic materials, the system had to be designed, built, and fine-tuned to achieve the levels of uncertainty that are required for the success of the measurements presented here.

The measurement methodology requires two steps. First, the coefficient of thermal reflectance must be determined for each of the surface materials to be scanned (calibration). Second, the changes in the surface reflectivity as a function of changes in temperature are measured with submicron spatial resolution using either a 16-bit CCD camera (new) or a CW laser (existing).

A schematic of the TRTG designed and built at SMU is shown in Fig. 1. The system combines the two different techniques for acquiring the reflectivity change induced by the temperature change on the surface of the activated device, namely, (i) a multipoint CCD camera approach, and (ii) a single-point laser-based approach.

*CCD camera-based TRTG:* In the case of the multipoint approach, the change of reflectivity is captured as the change in the intensity of the reflected light on each element (pixel) of a CCD camera. The advantage of the approach is that it is simpler to use, easier to vary the wavelength of the probing light (to maximize the $C_{TR}$ coefficient), has excellent spatial resolution (as low as 200nm) and is orders of magnitude faster than the single-point approach. Current limitations include that it cannot capture fast transient processes. In order to capture fast temperature transient with good measurement accuracy, the laser approach is used.

*Laser-based TRTG:* The probing laser beam is projected perpendicularly to the heated surface of the device under test (DUT) from which it reflects back along the optical path to the sensitive area of a photodiode. The intensity of the reflected light depends on the reflectivity (temperature) of the sample's surface. To overcome the inherently low signal to noise ratio, the activation voltage of the device is modulated, resulting in a modulated photodetector signal that can more easily yield the useful signal from the raw photodetector signal output. The photodetector signal, containing the change in surface reflectivity caused by the temperature variations of the DUT, is acquired with a lock-in amplifier (or an oscilloscope) and is then scaled according to the calibrated data. The limitations of using a lock-in amplifier or an oscilloscope were previously discussed [16]. The key difference between the two approaches is that the lock-in approach cannot be used to measure transient temperate fields, while the oscilloscope technique makes it possible to measure transient temperature with microsecond or better temporal resolution. However, the oscilloscope technique is less accurate than the lock-in technique. The temperature field over a desired area of an active device can still be mapped by repeating the procedure at multiple physical locations, which is achieved by precisely moving the probing head with submicron resolution.

As mentioned above, the calibration approach consists of determining the relationship between changes in reflectance and surface temperature. The change in reflectance is measured by a differential scheme involving two identical PDs in order to minimize the influence of fluctuations in the energy output of the probing laser. The sample temperature is controlled by a thermoelectric (TE) element and measured with a K-type thermocouple. It is worth noting that the calibration must be performed for each of the materials on the surface of each device where a mapping of the temperature is carried out.

## 3. RESULTS

To demonstrate the concept, the TRTG described above was used to acquire the surface temperature of a simple activated device. The device chosen was a plain gold microresistor, shown schematically in Fig. 2 and pictorially in Fig. 3. A series of micro-resistor devices were fabricated with widths of 4, 14, 26, and 50 μm, and lengths of 100, 200, and 500 μm.

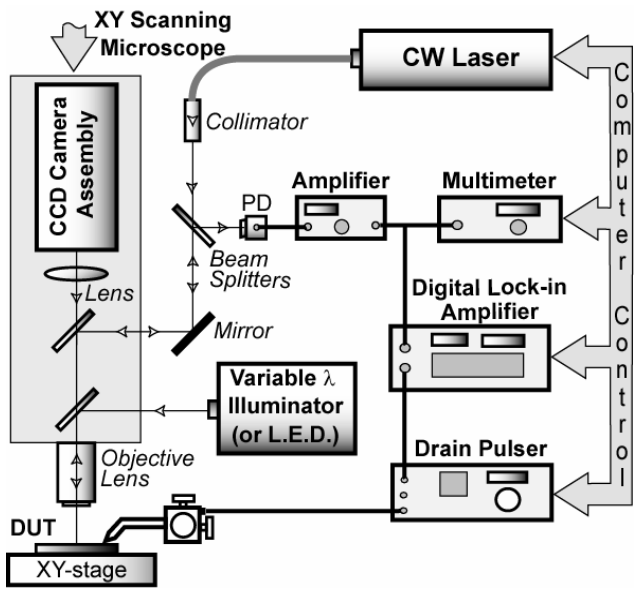

**Fig. 1** Surface temperature mapping system: combined laser (single-point) and CCD camera approaches

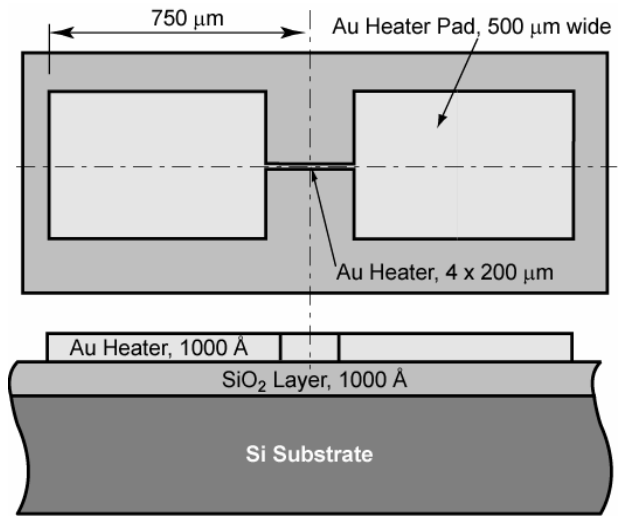

**Fig. 2 Geometry of the scanned micro-resistor**

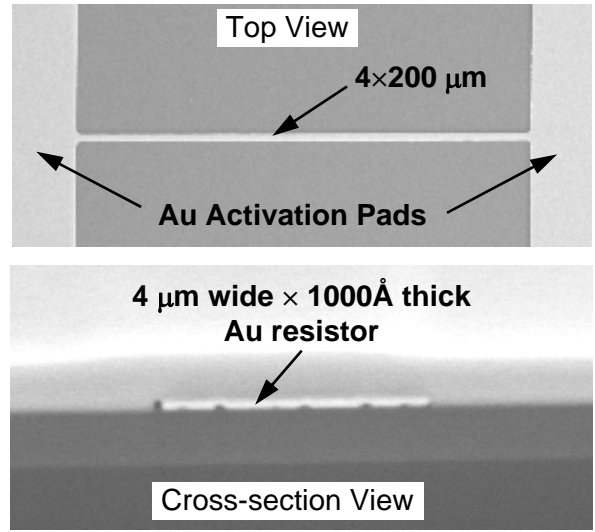

**Fig. 3 Top and cross-section views of the DUT: images taken using a Focus Ion Beam system**

The top and cross-sectional images of the micro-resistor shown in Fig 3 were obtained by the use of a Focus Ion Beam (FEI FIB-205) apparatus. A layer of platinum was deposited before etching away a section of the DUT using the same FIB machine, which was then used to view the cross-section of the DUT. The layer of Pt was necessary to protect the Au resistor from being damaged during the etching of the hole. The resulting hole was used to produce to view of the cross-section of the DUT in Fig. 3. An Electron Microprobe (JEOL JXA-733 Superprobe) was also used to image and study the geometry and chemical composition of the SUT.

A picture of the device and an example of the temperature contours measured with the TRTG system

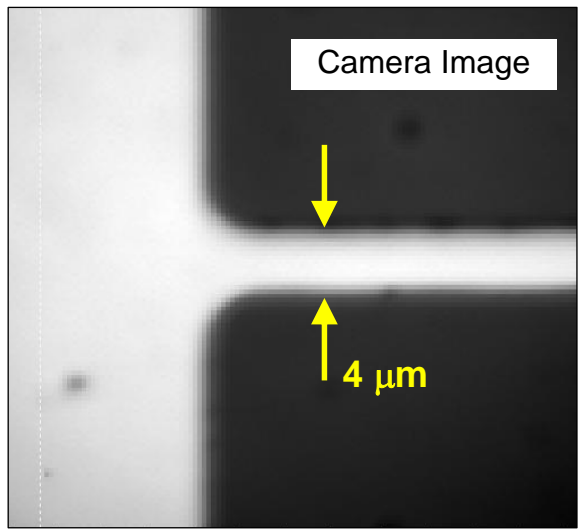

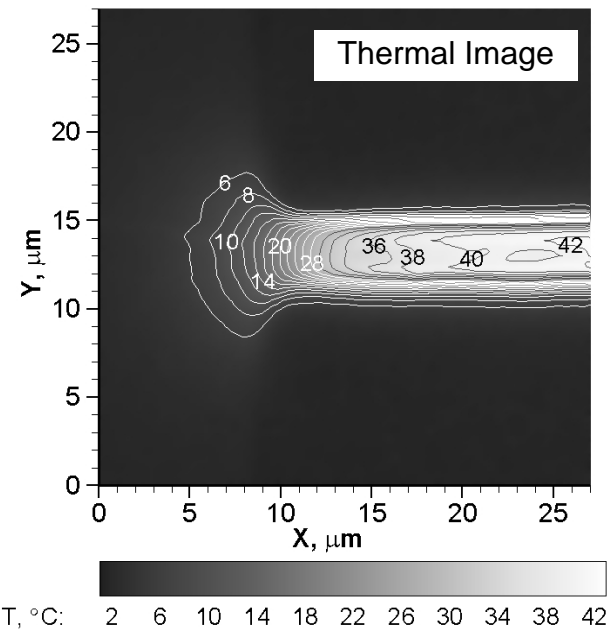

**Fig. 4 Scanned area of device and corresponding temperature contours on activated gold resistor device at instant of peak temperature: Illumination at 0.485 μm; Magnification = 75X; Spot size = 0.21 μm; Power = 97 mW**

*Pavel L. Komarov, Mihai G. Burzo, and Peter E. Raad*
*CCD Thermoreflectance Thermography: Methodology and Experimental Validation*

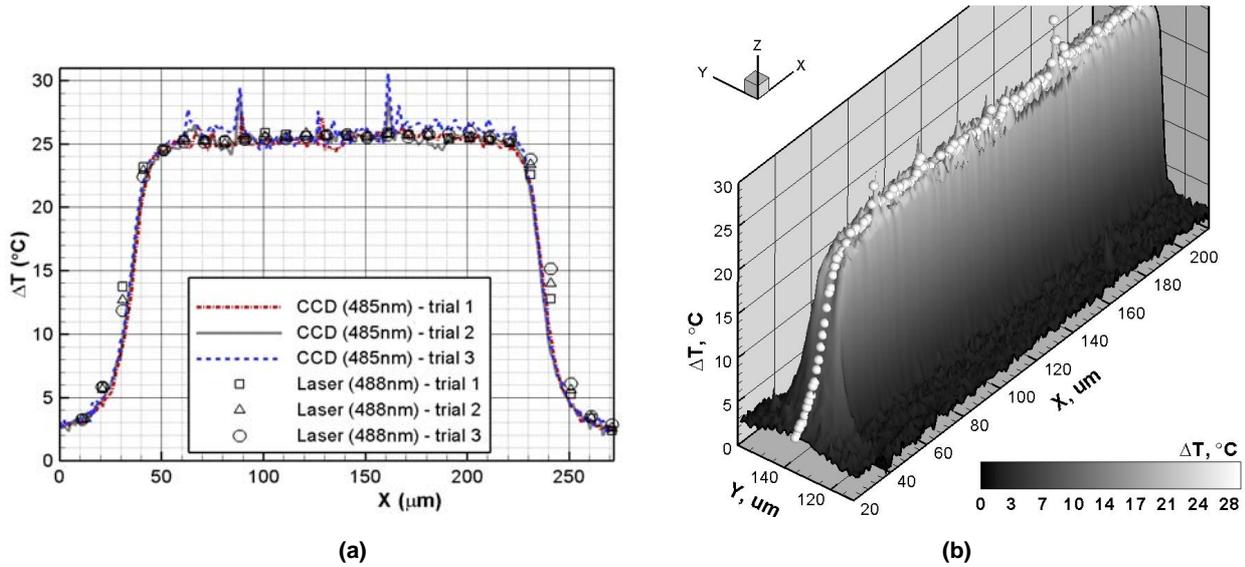

(a)          (b)

**Fig. 5** Comparison between measurements taken on 14x200 μm micro-heater using CCD and laser-based approaches: (a) Cuts along centerline of heater strip; (b) 3D elevation plot for CCD *trial 3* with laser data represented by white balls

(using the CCD camera approach) are shown on the left and right side of Fig. 4, respectively.

In order to validate the experimental temperature measurements obtained with the new CCD-based thermography system we compared the results with the results obtained using the existing single-point laser approach. A laser scan was performed along the centerline of the device using an Ar-Ion laser with a 488nm wavelength and a spot size of a few microns. The CCD measurements were carried out using the illumination from a blue (485nm wavelength) LED. A 20X objective lens was used for both measurements and the activation frequency was kept low (few Hz) to make sure that the steady-state temperature of the activated devices is captured. Figure 5 shows the comparison between the obtained results. The "spikes" are caused by dirt/defects on the surface of the sample that were picked up by the CCD camera.

The main reason for choosing to test a simple resistor is the fact that this enables us to validate the method by measuring its temperature with a different, completely independent approach. The approach chosen was the thermo-electrical method, in which the temperature change is determined by measuring the change in the electrical resistance of the resistor induced by Joule heating. A temperature coefficient of resistance of 0.0037 $K^{-1}$ [17] was used for the test resistor to obtain the absolute temperature change of the resistor and to compare it to the results obtained from the TRTG approach. The temperatures measured by the use of the thermoreflectance and thermo-resistance approaches are compared in Fig. 6(a), with visible excellent agreement between the results. The measured temperature field on the 4 μm by 100 μm device is shown on in Fig. 6(b). The minimum width of the resistor was limited by the capabilities of producing the mask and manufacturing the device at SMU, since the system can measures lines much smaller than one micron. It is worth pointing out that the duration of the whole measurement was only a few seconds.

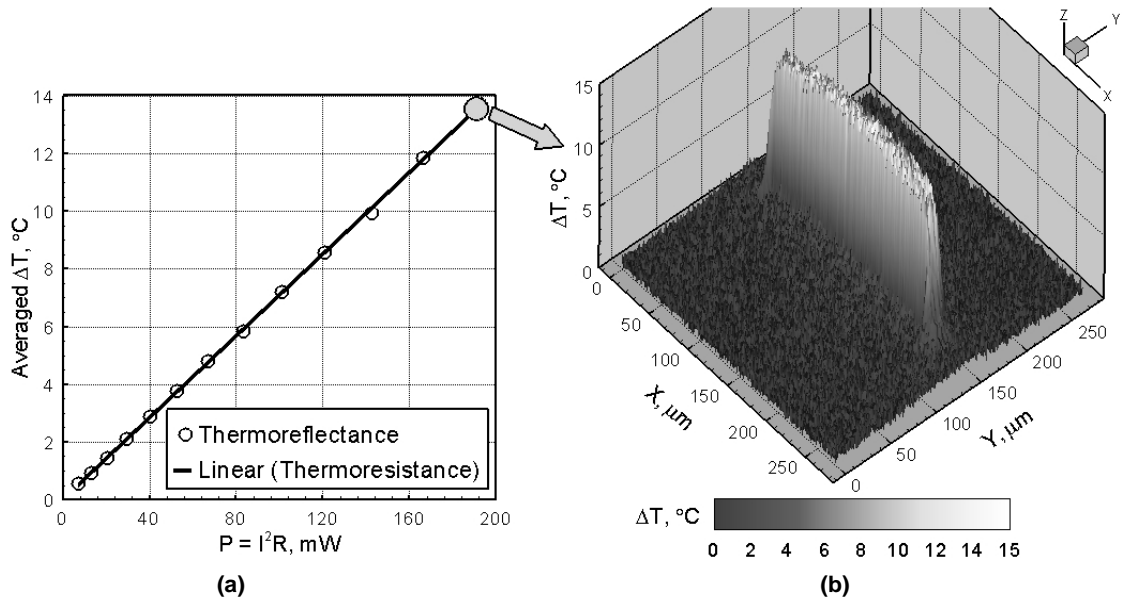

(a)          (b)

**Fig. 6** Measured average temperature change versus activation power for 14x200 μm resistor: (a) Comparison between thermoreflectance and thermo-electrical resistance measurements, (b) Measured temperature rise contours. (Thermoreflectance coefficient = 4.0 x $10^{-4}$ $K^{-1}$; Temperature coefficient of resistance = 0.0037 $K^{-1}$).

For existing devices, the highly resolved and accurate picture of the 2D temperature field would provide the ability to detect hot spots, diagnose performance, and assess reliability. In design and manufacturing of new devices, the new tool has the potential to provide a rapid approach for analyzing the thermal behavior of complex stacked structures, identify regions of excessive heat densities, and ultimately contribute to improved thermal designs, better device reliability, and shorter design cycle time. The outcome of this work will contribute to dealing with critical aspects facing the electronics industry and that are brought about by continued miniaturization, introduction of novel materials, and never-ceasing demand for higher performance.

## 4. CONCLUSIONS AND DISCUSSIONS

This work introduced a measurement system that is based on thermoreflectance physics and that uses a CCD camera to measure the surface temperature field of activated microelectronic devices with submicron spatial resolution. System details and outlines of the measurement methodology were presented along with a series of validation results. The results demonstrate that the newly built thermo-reflectance thermography system is capable of acquiring the surface temperature field of an activated DUT with up to 512×512 points and a spatial resolution of 0.2 μm. The measurement methodology and the features of the experimental setup were presented, along with details of the calibration process required to convert the changes in the measured surface reflectivity to absolute temperature values. The data acquisition procedure used to measure the transient temperature over a given active region of interest was also presented.

The system's capabilities were demonstrated on standard gold micro-resistors that were pulse activated and their surface temperature fields were measured at the instant when the device reaches its maximum temperature. A comparison between the results obtained by the use of the CCD camera approach and those obtained by our existing single-point, laser-based measurement approach was presented, with the data showing very good agreement (within 5%).

The system methodology was further validated by comparing the TRTG results with the data obtained from a non-optical approach. The independent temperature measurement technique consisted of determining the temperature of the micro-resistor by measuring the change in the resistance (using a 4-wire probe technique) induced by the Joule heating of the sample, and then converting that resistance change into a temperature change. The resulting temperature data was then compared to the averaged temperature of the gold heater.

## 5. ACKNOWLEDGEMENTS

The authors are grateful to Jay Kirk from the Electrical Engineering Department at SMU for his help in designing and building the devices presented in this work. We also thank Roy Beavers and Omniprobe Inc. for their help in testing the composition and geometry of the devices.